\documentclass[referee,useAMS,usenatbib,usegraphicx]{mn2e}

\title[ ATCA observations of the very young Planetary Nebula SAO 244567]{ ATCA\thanks{The Australia Telescope Compact Array is part of the Australian Telescope which is funded by the Commonwealth of Australia for operation as a National Facility managed by CSIRO} observations of the very young Planetary Nebula SAO 244567}
\author[G. Umana, C. Trigilio, L. Cerrigone, C.S. Buemi and P. Leto]{G. Umana$^{1}$\thanks{E-mail:
Grazia.Umana@oact.inaf.it} C. Trigilio$^{1}$, L. Cerrigone$^{2}$
\thanks{SAO predoc fellow, Harvard-Smithsonian Center for Astrophysics, Cambridge, MA 02138, USA},
C. S. Buemi$^{1}$, P. Leto$^{3}$ \\
$^{1}$ INAF- Osservatorio Astrofisico di Catania, Via S.Sofia 78, Catania, Italy\\
$^{2}$ Universit\'a di Catania, Dipartimento di Fisica e Astronomia, Via S. Sofia 78, Catania, Italy\\
$^{3}$ INAF-IRA, Noto, C.P. 161, Noto(SR), Italy.
}
\begin{document}
\date{}
%\date{Accepted 1988 December 15. Received 1988 December 14; in original form 1988 October 11}

\pagerange{\pageref{firstpage}--\pageref{lastpage}} \pubyear{2002}

\maketitle

\label{firstpage}

\begin{abstract}
The radio emission from the youngest known Planetary nebula, SAO\,244567,
has been mapped at 1384, 2368, 4800, 8640, 16832 and 18752 MHz by using the Australian
Telescope Compact Array (ATCA).
These observations constitute the first detailed radio study  of this very
interesting object, as they allow us to obtain the overall
radio morphology of the source and to compute, for the first time, the radio spectrum up
to millimetre range.
Radio emission is consistent with  free-free from a wind-like shell, which is also the region where
most of the [OIII] comes from as revealed by HST images.
Physical parameters of the radio nebula and of the central star were  derived, all consistent with SAO~244567 being a very young
Planetary Nebula still embedded in the dusty remnant of the AGB phase.
The optically thin radio flux density appear to decrease when compared to data from the literature. Even  very appealing, the variability of the radio emission, probably related to the evolution of the central object, needs further investigations.  

\end{abstract}


\section{Introduction}
The fate of a star with a main sequence mass in the range from  1 to 8 solar mass is well established: it  goes
through the asymptotic Giant Branch (AGB) phase, then into the Planetary Nebula phase (PN hereafter) and eventually
it will  finish its  evolution as a White Dwarf.
However, the details of these evolutionary phases, such as the formation and the early evolution of PNe  is far to be completely  understood. It is not clear, for example, which kind of mechanism changes the more or less regular and symmetric circumstellar envelopes
(CSEs) around PNe progenitors (AGB stars) into the quite complex morphologies observed in young and more evolved PNe.

New clues on the process of PNe formation can be provided by the analysis of
the physical characteristics of objects in the
short phase between the end of the AGB
and the onset of the ionization in the nebula. For this purpose, many authors
have tried to identify very young PNe or proto Planetary Nebula (PPNe) but this
has revealed to be quite difficult as this evolutionary phase is very rapid and because  the central object is
often heavily obscured by the thick circumstellar
envelope formed during the AGB phase.

Among post-AGB objects, SAO 244567 appears to be   unique, as it is evolved so rapidly that strong spectral
and total luminosity changes were followed
in a human life time scale.
\citet{b8} classified  SAO 244567  as a post-AGB star on the basis of its high galactic latitude and because of its far-infrared
(IRAS) colours similar to those of known PNe.
Very soon it was realized that,  in few decades,  its optical and ultraviolet spectra have developed
characteristics typical of the presence of a nebula making SAO 244567 one of the youngest PN never
discovered \citep{b9}.

In particular, its  optical spectrum has evolved quite rapidly: reported
by \citet{b5} with only $H_{\alpha}$ in emission,
more recently, it has shown strong forbidden nebular emission lines, which are  consistent with a young PN
\citep{b10}.
These results confirm that SAO 244567 has turned into a PN within the last 20 years and makes the
source a perfect target for studying  the early structure
and evolution of Planetary Nebulae.

SAO~244567 has been observed in 1996 with WFPC2 on board of the Hubble Space Telescope \citep{b3}.  
These observations pointed out that most of the nebular emission originates from a ring/ellipse, whose major axes extends for 
$ \sim  1.6^{ \prime\prime}$, and  the presence of low density collimated outflows. The authors interpreted the ring structure as the remnant of the mass-loss during the AGB phase, while the low-density bipolar structures as results from fast-wind experienced by the star during the
post-AGB phase. The existence of a companion star, detected  at
0.4 arcsec from the central star, indicates that the ring/eclipse is probably a circumbinary envelope and  that binarity  may play an important role in the shaping of PNe. These images are one of the best examples where  nebular structures,
that appear to collimate fast outflows, are evident.

As typical for very young Planetary Nebulae (YPN), SAO~244567 shows a strong infrared excess and it is
associated to the IRAS source IRAS $17119-5926$.
Measured non-corrected IRAS fluxes are 0.65, 15.50, 8.20 and 3.52 Jy
at 12, 25, 60 and 100 ${\mathrm \mu \mathrm m}$
respectively.

In spite of the numerous optical studies of this interesting object, very little is still known on its radio properties.
\citet{b9}  briefly reported on ATCA 6 and 3 cm observations
obtained in 1991. The measured flux densities are $ 63.3 \pm 1.8$ mJy and $51 \pm 12 $ mJy at 6 and 3 cm
respectively.

In this paper we present new ATCA observations of SAO 2444567 aimed to
determine the radio properties  of the nebula associated to this
very young PN. 

\section{Observations}
We carried out radio observations of SAO~244567 in March 2000 and August 2002.
In both epoch the  target was observed with the Australia Telescope Compact Array (ATCA) at Narrabri.
The ATCA consists of six 22-m antenna,
five of which lie along a three kilometer railway track, oriented east-west,
and the other antenna lies on a similar railway track three kilometers away thus providing a 6 km maximum
baseline lengths.

As each ATCA antenna is equipped with dual-frequency systems, it is possible to observe simultaneously at two different frequency bands.
%%%%%%%%%%%%%%%%%%%%%%%%%Figura 1%%%%%%%%%%%%%%%%%%%%%%%%%%%%%%%%%%%%%%
\begin{figure}
\resizebox{\hsize}{!}{\includegraphics{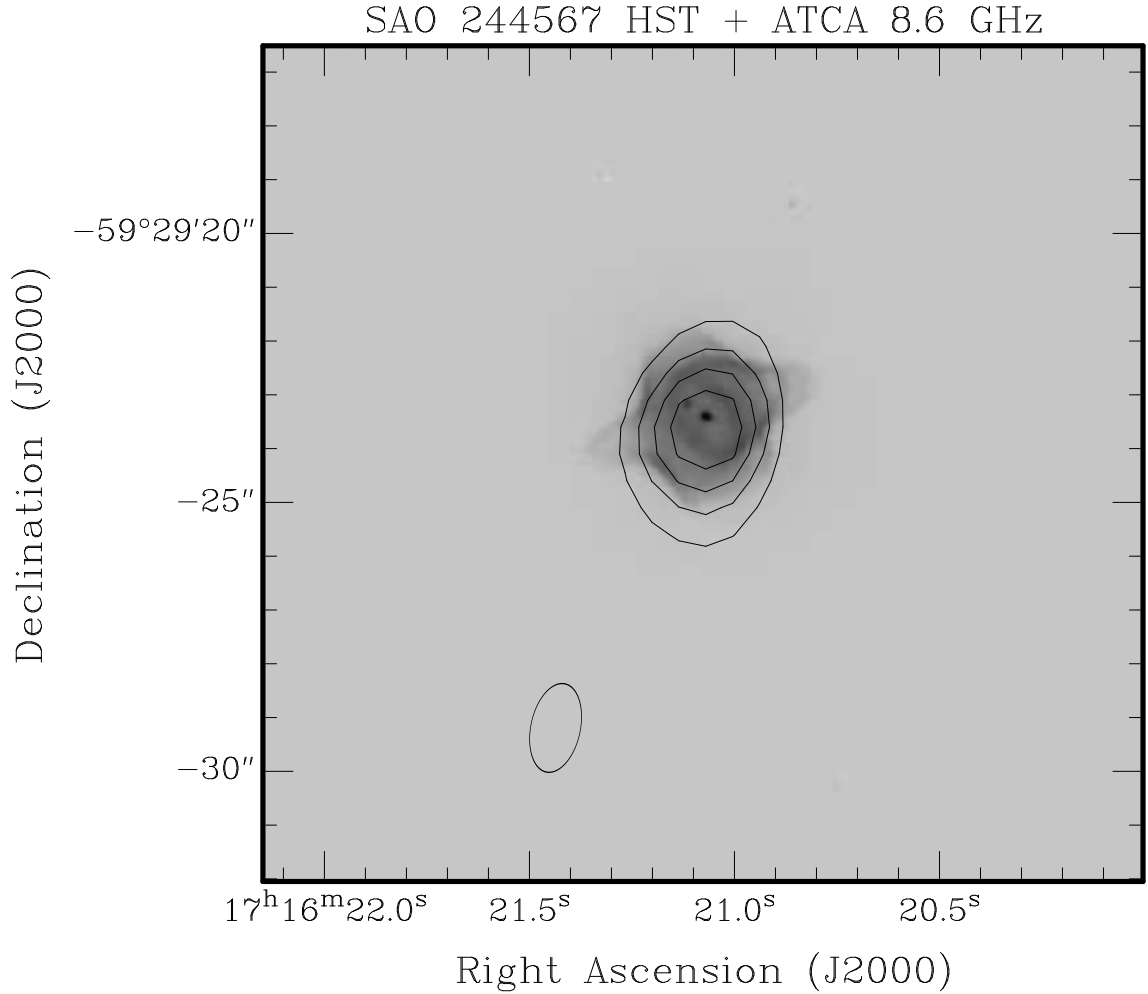}}
\caption{The radio map of SAO~244567 obtained with the ATCA at 8.4 GHz
(contour levels) superimposed with the HST image in $H_{\alpha}$
\citep{b3}. The synthetic beam is shown in the lower left corner.}
\label{map}
\end{figure}
%%%%%%%%%%%%%%%%%%%%%%%%%%%%%%%%%%%%%%%%%%%%%%%%%%%%%%%%%%%%%%%%%%%%%%%
\begin{table}
\caption{Summary of results: flux densities are determined by fitting of visibility curves.}
\begin{tabular}{lccccc}
\hline
{\em Frequency} & {\em Flux}         & rms            &  {\em phase cal Flux}\\
MHz             &  mJy               & mJy             & mJy \\
                &                    &                           &\\

\em{Epoch 2000}            &     &  &  \\
\hline

       4800          & $57.6  \pm 1.7$  & 0.3   & $4500 \pm 50$  \\
       8640          & $52.0   \pm 1.6 $ & 0.4   & $3400 \pm 50$  \\
                &                                 &                &\\
\em{Epoch 2002}             &     &   & \\
\hline
       1384          & $36.6  \pm 1.1$  & 0.2    &\\
       2368          & $46.9  \pm 1.8$  & 1.2    & \\
       4800          & $48.8  \pm 1.5$  & 0.2    & $5234\pm 20$   \\
       8640          & $46.6  \pm 1.4$  & 0.2    & $4271 \pm 20$ \\
      16832          & $43.8  \pm 2.0$  & 0.3    &              \\
      18752          & $42.8  \pm 2.0 $ & 0.3    &     \\

\hline
\end{tabular}
\end{table}
\subsection{The 2000 data}
In the  first epoch, March 2000, the observations were simultaneously performed at 6 (4800 MHz) and 3~cm (8640 MHz), with a 128 MHz bandwidth divided in 32 spectral channels, with an integration time of 30 sec. The observations were carried out in the
6D configuration, which, with a full synthesis,  usually provides a typical beam size ($\theta_{\mathrm {syn}}$)  of about
2$\arcsec$ at 6~cm.

The flux scale was fixed by daily observations of the primary flux calibrator
$1934-638$, whose assumed flux densities are 5.83 and 2.84 Jy at 4800 and 8640
MHz respectively.
In order to achieve a good phase calibration, the phase calibrator $1718-649$, which is 5.5 degree away from SAO 244567, was observed
at the beginning and at the end of each source observations.

At this  epoch the source was observed in the contest of a survey program aimed to detect radio emission from a sample of post-AGB stars. Therefore, for a good compromise between UV coverage and total integration time, the observations were performed in
snap-mode, which consists of short on source  runs carried out at
different hour angles.  A typical observation of our target consisted
of 7/8 cuts, each of 15 minutes long.
In the case of SAO 244567 we observed the source with  14 cuts, for a total  of 3.5 hours.
The observations were, however, obtained in different days starting from March 19 to March 22.

Due to serious  hardware problems at one of the antennas,  the observations suffered of a degradation of the available angular resolution, resulting in a beam of $ 7.6^{\prime \prime} \times 2.7^{\prime \prime}$ at 6 cm definitively worst than the
angular resolution  achievable with the 6D configuration.

\subsection{The 2002 data}
We have successively re-observed the source this time with a proper full tracking for better UV coverage.
The observations were carried out on August 24 and August 29, 2002. In  both dates the array was in 6C configuration. 
On August 24 we observe simultaneously at  16832 and 18752 MHz, with a 128 MHz bandwidth divided in 32 spectral channels, using an integration time of 30 sec. At that time only 3 antenna were equipped with mm backends. 
High frequency observations are strongly affected by atmospheric phase
fluctuations. To minimize this effect, we quickly switched between target and phase calibrator, with typical duty cycle of 6 min (four min on target and 2 min on phase calibrator) for a total of on source time of 7 hours. Reference pointing on our phase calibrator
($1718-649$) was also performed every hour of observing time.
Flux scale was fixed by several observations of $1924-638$ and the flux of $1718-649$ was
boot-strapped using only those scan of primary calibrator performed
at the same elevation as $1718-649$.

On August 29, the observations were carried out at 4800/8640 MHz and at 1384/2368 MHz. The observations were performed following the same procedure as
on August 24, using $1718-649$ as phase calibrator, with a typical on source scan of 15 minutes.
However, only $30\%$ of the total on source time of 10 hours was spent  observing at the lowest frequencies. The flux scale was fixed
relatively to $1934-638$ whose flux is assumed to be 14.94, 11.59, 5.83 and 2.84 Jy at 1384, 2368, 4800 and 8640 MHz.
%%%%%%%%%%%%%%%%%%%%%%%%%Figura 2%%%%%%%%%%%%%%%%%%%%%%%%%%%%%%%%%%%%%%
%\begin{figure}
%\resizebox{\hsize}{!}{\includegraphics{fit_spe.ps}}
%\caption{The observed radio spectrum of SAO244567 fitted with a shell, with
%decreasing density ($\propto r^{-2}$), with $T \sim 10^{4} K$ and $N_{e} \sim
%10^{4} cm^{-3}$. }
%\label{excess}
%\end{figure}
%%%%%%%%%%%%%%%%%%%%%%%%%%%%%%%%%%%%%%%%%%%%%%%%%%%%%%%%%%%%%%%%%%%%%%%
\section{Results}
The data from both epochs were reduced with the MIRIAD package, following the standard reduction steps.
Visibilities were weighted by applying the robust weighting (robust parameter=0.5).
This methods allows to obtain the same sensitivity as with natural weight
but with a much better beam shape.
Finally the obtained maps were cleaned down to few times the theoretical noise, estimated to be of the order of 0.1--0.2 mJy.

The map with best combination of resolution and sensitivity is that obtained at 8640\,MHz in 2002.
The radio map (Fig.~\ref{map}) reveals a slightly extended structure,
whose overall size
($\sim 1.7^{\prime\prime}$) compares quite well with
that observed in H$\alpha$ \citep{b3}. 
However, the limited ATCA angular resolution prevented us to point out any fine detail of the radio structure.

Total flux density and the angular dimension  can be usually estimated directly  from the obtained map, by a 2D Gaussian fit of the radio source. However, because an interferometer will spatially filters out structures larger than the resolution corresponding to the minimum antenna spacing, the evaluation of flux density by Gaussian fit becomes less precise for partially resolved sources. 
This effect will be more evident at higher frequencies. In this cases it is advisable to derive flux densities directly from UV data, by fitting the visibilities  at zero baseline. 
When the sources are angularly small, different methods to derive the flux density, i.e. from the map or from the fit of visibility data, provide consistent flux values to within a few $\%$. 
 In our case, differences between total fluxes as measured from the map and from fit at zero baseline are of the order of 0.1
percent up to 4800 MHz, but become more important at 18750 MHz where it goes up to $\sim 17\%$.
For this reason, in the following analysis we will considered the flux density derived from visibility fits.   
Results are summarized in Table~I, where the observing frequency, the measured flux density, with its associated $\sigma$ and  the rms of the visibility fit are reported. An error in the flux calibration can result in a  systematic error 
in flux of the observed source.
Therefore the $\sigma$ associated to the flux density estimation is derived from:
%\begin{equation}
\[
  \sigma = \sqrt{ (rms)^{2} + (\sigma_{\mathrm {cal}} S)^{2}}
\]
%\label{eq2}
%\end{equation}
where rms is the error associated to the fit,  $\sigma_\mathrm{cal}$ is the error associated to the absolute flux density scale, which is accurate to within few (3--5) percents  and $S$ is the derived flux density.\\
In order to successively compare results from different epochs, for frequencies with multi-epoch data  we also report, in the last column of Table~I,
the flux density of the phase calibrator ($1718-649$).

We note that the flux density at 4800 and  8640 MHz appears to decrease from 1991 \citep{b9} to 2002 (this paper).
This decrement appears to be real and not related to a variation in the absolute flux scale as the flux density of the phase calibrator changes between the last two epochs but with a different trend, and it is actually increasing maintaining the same spectral index.
In the following analysis only the multi-frequency epoch 2 data set will be used.

\begin{table}
\caption{Parameters of different models for the radio nebula around SAO 244567.}
 \begin{tabular}{@{}lr@{}}
Parameter & \multicolumn{1}{c}{Value} \\
                                & \\
{\bf Model 1 (Wind  Shell)}		&\\		
\hline
Density			        & $ n(R_{\mathrm {int}})= 2.5 \times 10^{4} {\mathrm {cm}}^{-3} , ~~n\propto  R^{-2}$    \\
Shell internal radius $R_{\mathrm {int}}$	& $0.75$ arcsec \\
Shell external radius $R_{\mathrm {ext}}$                 & $1.5$ arcsec \\
Shell Temperature               & $ 9600$ K\\
\noalign{\smallskip}
                                & \\
{\bf Model 2 (Shallow Shell)}		&\\		
\hline
Density			        & $ n= 1.45 \times 10^{4}$   ${\mathrm {cm}}^{-3}$ \\
Shell internal radius			& $0.65$ arcsec \\
Shell external radius                 & $1.3$ arcsec \\
Shell Temperature               & $ 10000$ K\\
\noalign{\smallskip}
                               & \\
{\bf Model 3 (Sphere)}		&\\		
\hline
Density			        & $ n= 1.23 \times 10^{4}$   ${\mathrm {cm}}^{-3}$ \\
Sphere size			& $1.4$ arcsec \\
Sphere Temperature               & $ 10000$ K\\
\noalign{\smallskip}
\end{tabular}
\end{table}

\section{The Nebula}
\subsection{The Radio Morphology}
\begin{figure}
\resizebox{\hsize}{!}{\includegraphics{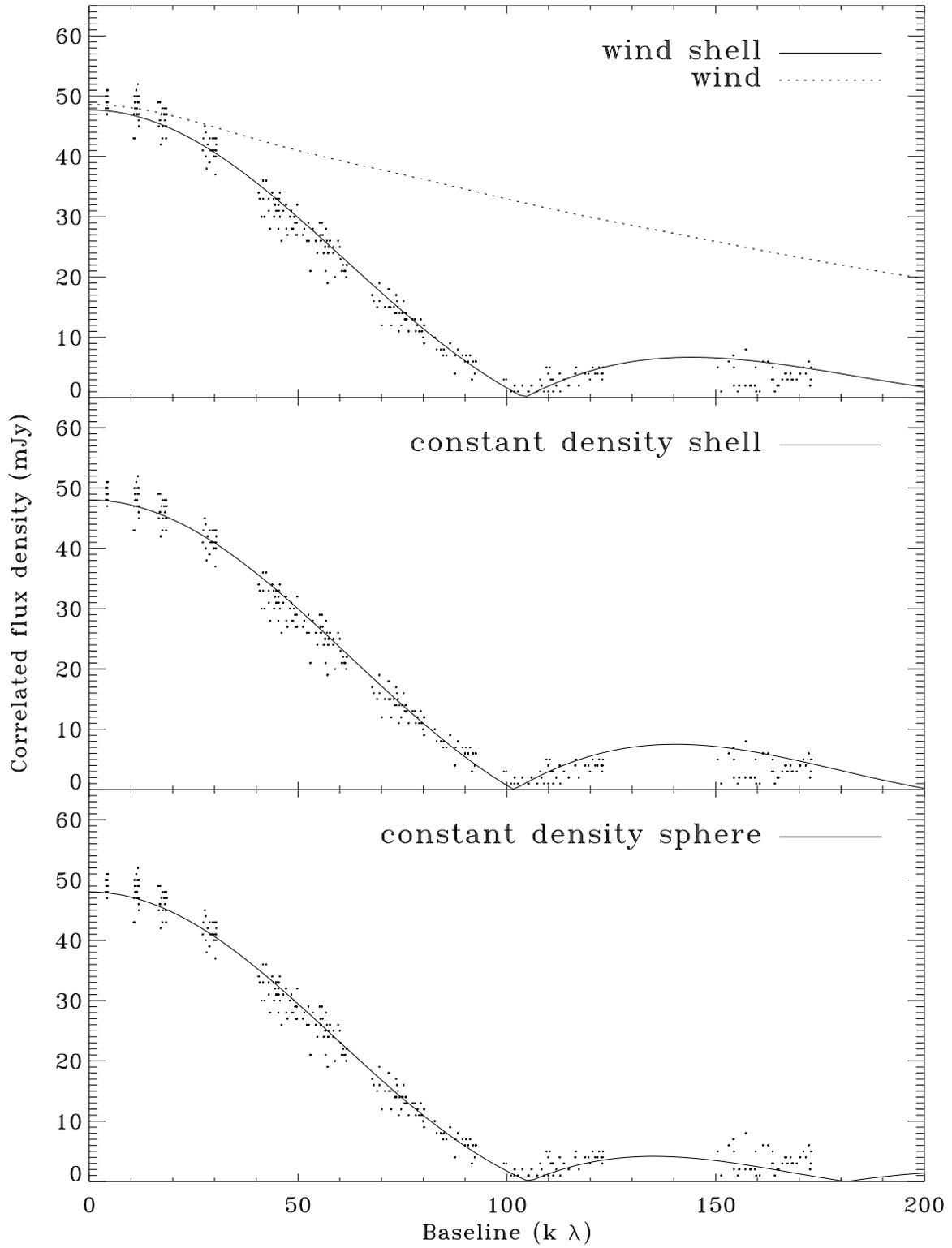}}
\caption{Comparison between different modelling of the radio source.
Both  a shallow shell (with density $\propto \frac{1}{r^{2}}$ or constant) and a
constant density sphere are consistent with the observed visibilities.
A stellar wind extending to infinity is shown as example.}
\label{models}
\end{figure}

Some information on source morphology can be derived from the analysis of the visibility 
data, i.e. the fringe amplitude as function of the interferometer spacing.
Since  the ATCA is a linear interferometer, each of the different 15 minutes on source scans (cuts)   
will give a maximum resolution in one
direction and the corresponding visibility will be function of the source morphology in that
direction. Different visibility corresponding to different cuts look quite similar and this
indicates that the overall morphology of the radio source is symmetric with respect to a central point.
The small   elongation  in the S-W direction, probably reflects the UV coverage, and therefore the synthetic beam with which the visibility data are convolved and it is not intrinsic to the source.  \\
We have modelled the visibilities of the entire observing run, averaged over 15 minutes,
assuming 3 different symmetric morphology for the radio source, i.e.: 1) a shallow shell,
$0.75 ^{\prime\prime}$ thick, with a density decreasing with distance from the central star ($n \propto \frac{1}{r^{2}}$); 2) a constant density shallow shell, $0.65 ^{\prime\prime}$ thick; 3) a constant density
sphere, with a radius of $1.4 ^{\prime\prime}$.\\
The best-fit parameters of the 3 models are summarized in Table II and results, relative to visibilities observed at 8640 MHz, are shown in Fig.~\ref{models}. In the first panel the expected visibility from a stellar wind is shown (dotted line) for comparison.
From this analysis we may conclude that the source morphology is consistent with both a shallow shell (with constant
or with  $\propto \frac{1}{r^{2}}$ density) and with a constant density sphere. 

%%%%%%%%%%%%%%%%%%%%%%%%%Figura 2%%%%%%%%%%%%%%%%%%%%%%%%%%%%%%%%%%%%%%
%%%%%%%%%%%%%%%%%%%%%%%%%%%%%%%%%%%%%%%%%%%%%%%%%%%%%%%%%%%%%%%%%%%%%%%

%%%%%%%%%%%%%%%%%%%%%%%%%Figura 3%%%%%%%%%%%%%%%%%%%%%%%%%%%%%%%%%%%%%%
\begin{figure}
\resizebox{\hsize}{!}{\includegraphics{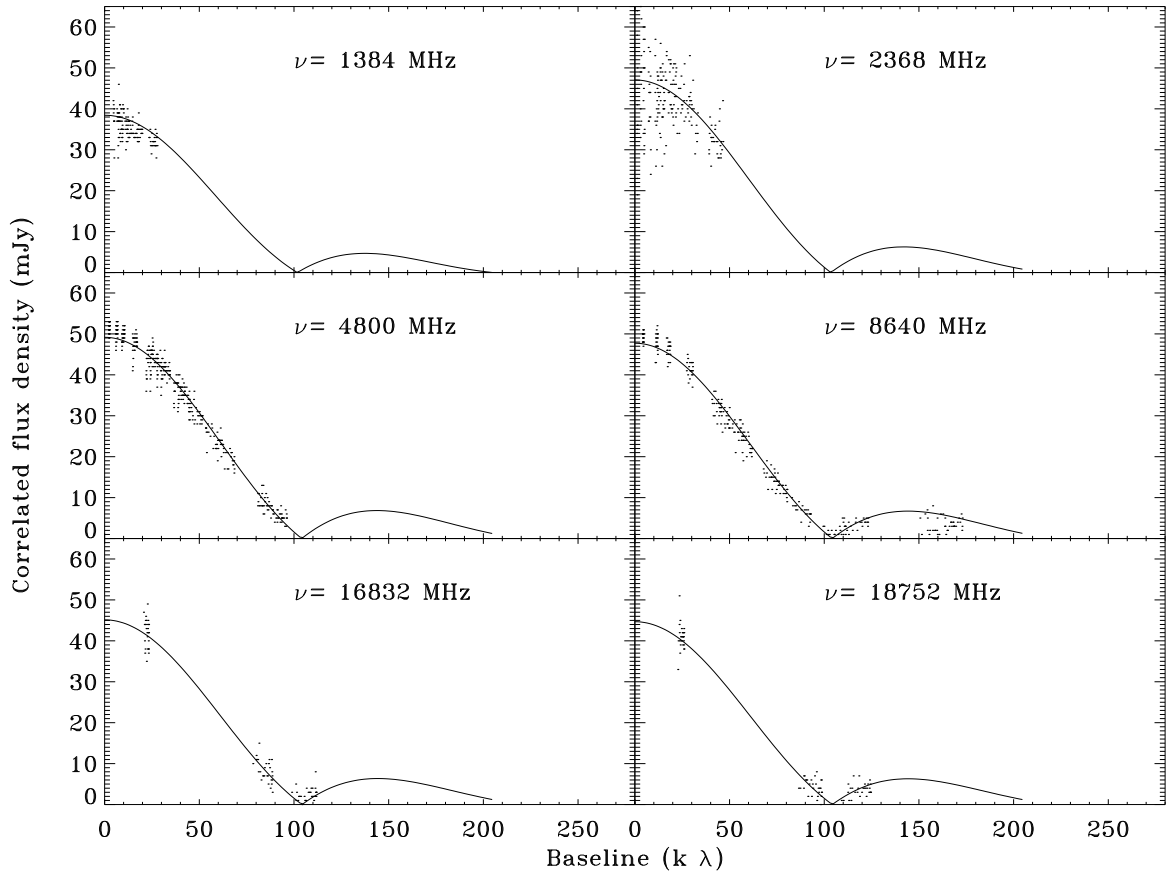}}
\caption{The modelling of visibilities at various frequencies.
Fits are obtained assuming a shallow shell, with
decreasing density ($\propto r^{-2}$), with $T=9600$ K and $n=2.5 \times 10^{4}$ cm$^{-3}$ at the inner radius}
\label{visibility}
\end{figure}
%%%%%%%%%%%%%%%%%%%%%%%%%%%%%%%%%%%%%%%%%%%%%%%%%%%%%%%%%%%%%%%%%%%%%%%

%%%%%%%%%%%%%%%%%%%%%%%%%Figura 4%%%%%%%%%%%%%%%%%%%%%%%%%%%%%%%%%%%%%%
\begin{figure}
\resizebox{\hsize}{!}{\includegraphics{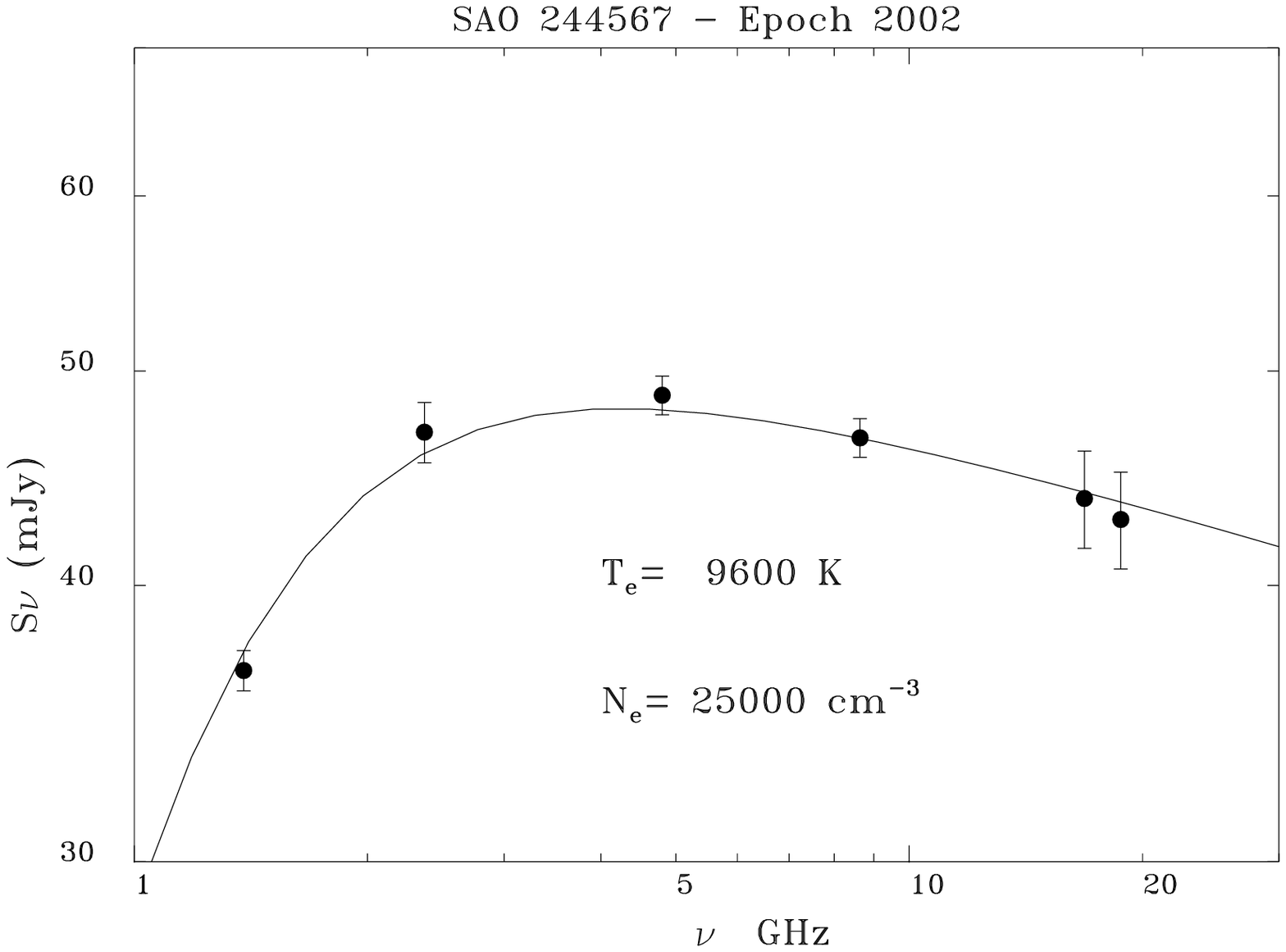}}
\caption{The radio spectrum of SAO~244567 obtained from the fit showed in figure
taking as flux density the zero spacing value.}
\label{spettro}
\end{figure}

\subsection{The physical properties}
Information on several parameters of a radio source can be derived by analyzing its radio spectrum, in particular when the 
transition region between optically thick and optically thin regime can be pointed out.
The analysis of visibility data indicates a symmetric morphology, consistent with both a constant density sphere and a wind or constant density shell. Very little studies  on modelling of free-free radio spectrum of PNe have been conducted, 
mostly because of the very few multi-frequency spectra available in the literature.
Among them, \citet{b1}, by analyzing a sample of compact (young) PNe, noted a trend of increasing value of spectral 
index $\alpha$ (flux $ \propto \nu^{\alpha}$), evaluated in the optically thick part of the spectrum, as the turnover frequency ($\nu_{\mathrm c}$) decreases. This was interpreted as evidence, in the very early stages of nebula evolution, for most of the free-free emission coming from the AGB progenitor's wind, with typical $\alpha=0.6$.  Taylor et al. (1987) modelled the radio continuum of 18 compact PNe with a shell model, with a radially dependent density.  Observed radio spectra are well represented by a wind-shell, supporting the idea that free-free originates from a
region formed with a constant stellar mass-loss during the precedent red giant phase.

These precedent results together with  the fact that SAO~244567 is a very young PN, will lead us to assume for SAO 244567 a wind-shell morphology. Model 1 well reproduces the observed visibilities at each frequency (Fig.~\ref{visibility}).
The foreseen radio spectrum, superimposed to the observed data, is shown in Fig.~\ref{spettro}. 
with a turnover frequency, i.e. the frequency at which the nebula becomes transparent, between 2368 and 4800 MHz.
The model will allow us to get an estimate of the total ionized mass contained in nebular shell. 
In the hypothesis of a pure hydrogen nebula at 5.6 Kpc \citep{b9}, we derive a total ionized mass of  $ \sim 0.065 ~M_{\sun}$. This value depends on the value of volume
filling factor of the nebula, which has been assumed to be 1. A similar value was derived by \citet{b9}, 
while \citet{b2} derived a value of $ 0.2 M_{\sun}$ from the $H_{\beta}$ flux.

In the spectral region where the source is optically thin we can use the observed radio flux density
plus the angular size to derive the mean emission measure as follows  \citep{b16}:
%\begin{equation}
\[
  <EM>=\frac{ \int_{\Omega} EM  d\Omega}{ \Omega} =\frac{ 5.3 \times 10^{5} F_\mathrm{8.4~GHz}}{ \theta^{2} }=1.1 \times 10^{7}
\]
%\label{eq7}
%\end{equation}
where $F_\mathrm{8.4~GHz}$ is the measured radio flux density, in mJy, at 8.4~GHz,
and $\theta$, in arcsec, the angular dimension of the
radio emitting region as derived from the model. The mean emission measure is expresses, as usual, in cm$^{-6}$pc.\\
Young PNe should have emission measures of the order of $10^{6}-10^{8}$cm$^{-6}$pc \citep{b16,b6}. 

From the radio optically thin flux we can also derive  
the excitation parameter ($U_{\rm exc}$)  required to account for the measured radio flux:
\begin{equation}
 U_{\rm exc} = 13.3 ( \nu^{0.1} T^{0.35} D_{\rm Kpc}^{2} F_{\nu})^{ \frac{1}{3} }
~~~\mathrm{pc~cm}^{-2}
\label{uexc}
\end{equation}
where $F_{\nu}$ is the optically thin radio flux density, expressed in Jy, at the observing frequency  $\nu$
(in GHz); $T$ is the nebula temperature, expressed in $10^{4} {\rm K}$, and $D_{\rm Kpc}$ is the source distance in Kpc.\\
As  the PN associated to SAO 244567 is very young we can assume that it is ionization bounded; in this case  the excitation parameter  is directly  related to the number of ionizing photons emitted by the central star.
\begin{equation}
  L_{\rm uv}= 1.23 \times 10^{56} \beta U_{\rm exc}^{3} ~~~\mathrm{photons~s}^{-1}
\label{luv}  
\end{equation}
where $\beta$ is the hydrogen recombination coefficient summed over all levels above the ground level
($\sim 3 \times 10^{-13} \mathrm{cm^{3}~s^{-1}}$ for
$ T \sim 10^{4} $ K)  \citep{b13}.\\
From equation~(\ref{uexc}) and equation~(\ref{luv}), assuming a nebular temperature of  $ T \sim 9600$ K
and a distance of 5.6 Kpc, from the radio flux density measured at 4.8 GHz, we derive a
number of ionizing photons of
  $L_{\rm uv} \sim  1.5 \times 10^{47} (\mathrm{photons~s}^{-1})$.\\
This corresponds to the flux of Lyman continuum photons of a B0--B0.5 V
with an effective temperature of $T_{\rm eff} \sim (2.8 \pm 0.2) \times 10^{4}$ K
\citep{b7}.\\
This value for the effective temperature of the central star is somehow lower than the value of $T_{\rm eff} \sim 3.7  \times 10^{4} {\rm K}$
estimated by \citet{b9} from the analysis of the UV spectra and corresponding to a O8V star.
This may indicate the presence of significant quantity of dust 
well mixed with the gas in the inner regions of the nebula. Dust, in fact, competes with the gas for the absorption of the ionizing photons and for a given number of ionizing photons emitted by the central star, it will reduce the observed radio flux.

\section{ The dusty envelope}
One of the most important results from PNe studies based on IRAS data was the realization that, on the average, 
in young PNe about $40\%$ of the emergent flux is emitted in the far-infrared \citep{b18}.
This is due to the presence of an extended dusty envelope, around YPNe, which is the remnant of the precursor's wind not yet dispersed.
In YPNe the spectrum emitted by the dust, between 10 and 100 $\mu {\mathrm m}$,  is usually well described by a single temperature black body curve \citep{b17, b14}.
%(Zhang \& Kwok, 1990; Stasinska \& Szcerba, 1999).  
We therefore derive a dust temperature of $T_{\mathrm {dust}} = 137 \pm 2$       
by using a least-squared fitting procedure to th IRAS measurements. This procedure provides, as a by-product, the total far-infrared
flux ($F_{\mathrm {IR}}$), obtained by integrating, over  the IRAS band (25 to 100 $\mu {\mathrm m}$), the  Planck curve  that fits  the IRAS data.

It is therefore possible to define the
far infrared excess (IRE)  as the ratio of the observed total far infrared flux
($F_\mathrm{IR}$) over the expected total infrared flux.
Under the hypothesis that the far infrared flux is due to thermal emission from dusty grains heated by
Ly$\alpha$ photons,
this ratio is unity. However, in young PNe heating by direct starlight is important and IRE can be
much higher than unity.
\citet{b20}
%Pottasch (\cite{Pottasch84a}) 
has derived a formula to express the expected total infrared flux
in terms of optically thin radio flux density:
%\begin{equation}
\[
  IRE=1.07 \frac{F_{\mathrm {IR}}}{F_\mathrm{8.6\,GHz}}=2.8 %FIR=125
\]
%\end{equation}
where $F_{IR}$ is expressed in $10^{-14}$ W\,m$^{-2}$,   $F_{8.6\,\mathrm {GHz}}$ in mJy, and the high density approximation has been adopted.
This value is consistent with the hypothesis that SAO~244567 is  a very YPN where the dust
plays an important role. \\
To quantify the total mass of the dust we will use the optically thin expression \citep{b21}:
%\begin{equation}
\[
  M_{\mathrm {dust}}=\frac{F_{60\mu \mathrm m} D^2}{\chi B\nu(T_{\mathrm {dust}})}
\]
%\end{equation}
where $F_{60\mu {\mathrm m}}$ is the flux density measured at $60 \mu {\mathrm m}$, 
D is the distance to the source (Kpc), $ B\nu(T_{\mathrm {dust}})$ is the value of the BB function of $T_{\mathrm {dust}}$, 
evaluated at $60 \mu {\mathrm m}$ and $\chi$ is the mean absorption coefficient for the grains (cm$^{2}$ g$^{-1}$). 
This formula provides only an estimation of the total dust mass as a single temperature and an unique size for the grains is assumed.
Moreover, while this dust mass determination is independent on the distribution of the grains, it strongly depends on the chemical composition of the grains. \\
There are still problems in understanding the chemical composition of the dust component in PNe. Usually, in carbon rich PN, dust is 
mainly composed of carbon-based grains, while in O-rich PN, dust is mainly composed of different forms of silicates.
As for SAO~244567 we  don't have any information on its chemical composition, we compute the dust mass for two different value of  $ \chi$, namely 53.45 cm$^{2}$ g$^{-1}$, which corresponds to circumstellar silicates and 145.32 cm$^{2}$ g$^{-1}$, which corresponds to graphite \citep{b14}.
This results in a total mass dust of $2 \times 10^{-4} M_{\odot}$, in case of silicates, and 
$7.5 \times 10^{-5} M_{\odot}$, in case of graphite.
If we assume that dust and gas occupy the same volume, we may derive a dust to gas ratio of $3 \times 10^{-3}$  and $ 1.15\times 10^{-3}$ for silicates
and graphite respectively. These values are in the range for %$\frac{M_{\mathrm {dust}}}{M_{\mathrm {gas}}}$ 
${M_{\mathrm {dust}}} \ {M_{\mathrm {gas}}}$ 
derived for a large sample of PNe \citep{b14}.

\section{The variability of radio emission}
\begin{figure}
\resizebox{\hsize}{!}{\includegraphics{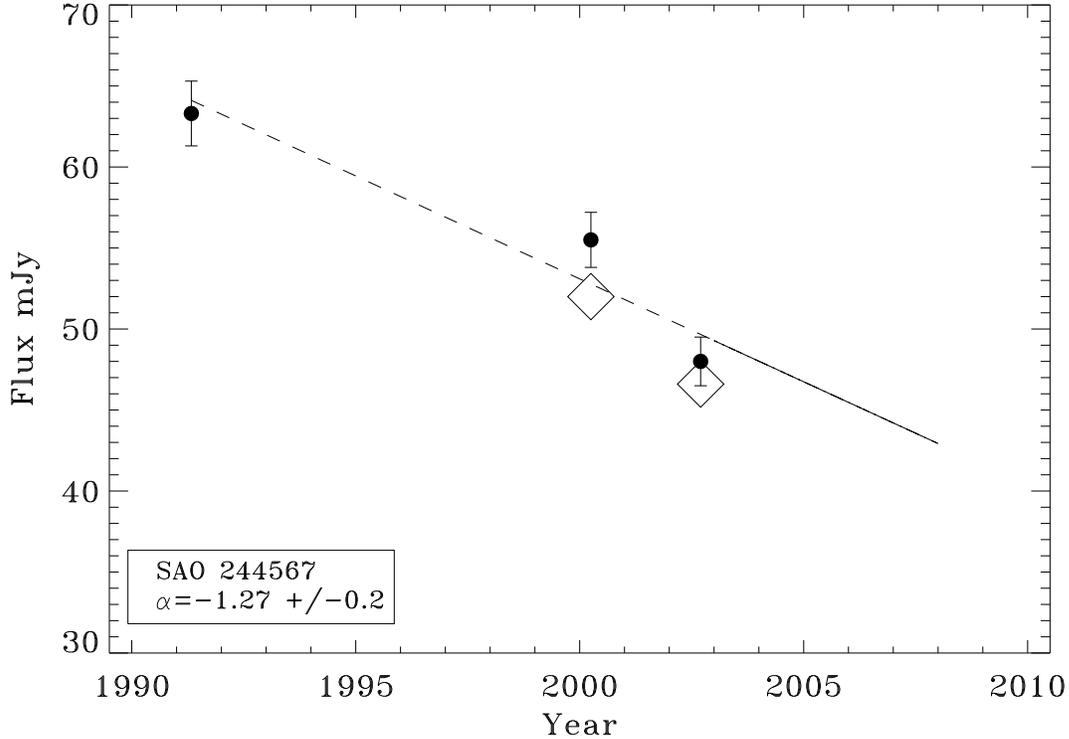}}
\caption{The behaviour of flux density of SAO 244567 as measured at 4800 MHz, filled dots.
Data are from \citet{b9} and this paper. A weighted fit to the data is also shown, which foresees a
radio flux density at this frequency of 43 mJy at the beginning of 2008. Measures at 8640 MHz are also
plotted, diamonds.}
\label{variability}
\end{figure}

One of the most striking and unexpected characteristic of radio emission from SAO 244567 is its variability.
In Fig.~\ref{variability}, we report  the flux density  at 4800 MHz, as measured by \citet{b9} and in this paper, indicating a linear decreasing trend of the optically thin flux. The dashed line is the weighted fit to the 4800 MHz data points, which results in a flux decrement of
$\sim 1.3$ mJy year$^{-1}$  
%~~ \frac{mJy}{year}$.  
Assuming  that the radio emission from the source will keep this trend a flux density of  $\sim 43$ mJy in 2008 is foreseen.
In the same figure we also include the 2000 and 2002 8640 MHz data, without considering the quite noisy datum from \citet{b9}.

 In the hypothesis that, between March 2000 and August 2002 ($\Delta T$), the ionized mass remains  the same (i.e.
 any effects due to the evolution of the central object are not considered), and the 
 nebula is expanding homologous, we have: 
 \begin{equation}
 M_{\mathrm {Tot}} \propto n_{2000}R_{2000}^{3}=n_{2002}R_{2002}^{3}
 \label{mtot}
 \end{equation}
 
  Starting from the results for epoch 2002, from equation~(\ref{mtot}) 
 we obtain constraints for epoch 2000.
 To reproduce the flux density observed at 4800 and 8640 MHz in epoch 2000, we then used  the model, as 
 illustrated in par 4.1, obtaining a reasonable fit that satisfies  equation~(\ref{mtot}), for  $T_{\mathrm {eff}}=9600$ K, as in 2002 fit,  with an  internal radius of $0.71^{ \prime\prime}$, an external radius of $1.4^{\prime\prime}$ and a density, at the base of the shell, of $n_{2000}=2.9~ 10^{4}$ cm$^{-3}$.
 To obtain such an expansion we  need the nebula to expand in a time $\Delta T$  with an average velocity of $ \sim 850$ km ~sec$^{-1}$. 
 We should stress here that a variation of radius of the order of few 0.1$^{ \prime\prime}$ would be not detectable with present
 instrumentation (ATCA). 
 
 High expansion velocity have been observed is the ultraviolet lines of some PPNe and YPNe \citep{b4} and are usually related  to fast-winds events occurring  during the post-AGB phase. For SAO 244567 \citet{b10} reported, 
in 1988, an ultraviolet spectrum characterized by strong P-Cyg profiles, indicating the presence of a strong wind with expansion velocities of the order of $\sim 3000$ km sec$^{-1}$. This spectral structures were observed until 1993 and by 1994 they were almost vanished, indicating the end of a rapid mass-loss event.
 At the same times, \citet{b9} derived,  from high-resolution [OIII] profile, an expansion velocity of the order of $\sim 8$ km sec$^{-1}$, consistent with the fact that the ionized shell, from which most of the [OIII] comes from, is the remnant of mass-loss during the previous AGB phase. 
 Both results are consistent if we assume that UV lines are produced in the low-density polar region, where episodic rapid mass-loss event 
 can occur,  while the bulk of the ionized material, and also the radio emission, is localized in the main ring/shell, as 
 indicate by the HST images, which is expanding with typical velocity as expected for a AGB wind.
 Therefore, we may conclude that an expansion of the main shell can not explain the observed decrement in the radio flux density because 
 the required expansion velocity is too high.
 
 High velocity mass-loss events may, however, contribute to the total ionization of the nebula.
 This is what is claimed by \citet{b11} to explain  the decrease in millimetic, free-free optically thin flux observed in the PPNe CRL~618.  According to these authors, in this source there is a substantial contribution of shocks, originating from interaction between the fast post-AGB wind with the CSE environment, to the total ionization. As the fast wind stops, the ionization rely only on the central star radiation field and the optically thin radio emission is, therefore, decreasing.
  However, this explanation requires  episodic high-velocity mass-loss events  to justify the flux increasing reported by the same authors 2 year later \citep{b12}.
A decrease in the optically flux density has been also noted in NGC~7027 \citep{b19} and it is 
interpreted as related to the evolution of the central object. NGC~7027 is located just at the tip of the WD cooling track,
 where its luminosity starts to decrease while its temperature is still increasing. This leads to a decreasing number of ionizing photons.
 Since the optically thin free-free emission is directly related to the ionizing flux from the central object, this would imply a decreasing optically thin radio flux density. \\
 SAO 244567 has shown many spectral changes in the last 20 years. Most notably, the changes in the UV flux level, which 
 suffered a factor 2.83  decrement in 7 years, while the central was gradually becoming hotter \citep{b10}, implying a total drop of the central star luminosity. This place the central star of SAO 244567 in the same position on the HR diagram as NGC~7027 \citep{b19}, just on the top of the cooling WD track. However, as already pointed out by \citet{b10, b3}, such a very fast evolution is very difficult to be explained in the framework of  actual post-AGB evolutionary models, that for a core mass object of $\sim 0.056 M_{\odot}$ \citep{b9}, foresee a much slower evolution.
  %%%%%%%%%%%%%%%%%%%%%%%%Figura 1%%%%%%%%%%%%%%%%%%%%%%%%%%%%%%%%%%%%%%
%%%%%%%%%%%%%%%%%%%%%%%%%%%%%%%%%%%%%%%%%%%%%%%%%%%%%%%%%%%%%%%%%%%%%%%
\section{Summary}
We have presented ATCA multi-epoch, multi-frequency observations of the YPNe 
SAO~244567, aimed to derive its radio characteristics to complete the picture
of this quite intriguing object.
The best radio map, for sensitivity and resolution, has been obtained  at 86400\,MHz (3.6 cm), 
which reveals a slight extended radio
structure. The angular resolution of ATCA does not allow to evidence the
fine details of the nebula as shown by HST observations.
However, some morphological information have been derived by a model to the observed visibilities data, which are consistent with 
a wind-like shell whose external radius is in agreement with that of the main structure observed by HST.

The mean emission measure and the infrared
excess, as derived by our observations, are consistent with a very
young planetary nebula, still embedded in its dusty envelope, remnant of the earlier AGB phase.
By comparison between total ionized gas mass, as derived from the radio, and the total dust mass, as derived
from the IRAS data, a dust to gas ratio has been derived, by considering two different types of chemistry for the 
dusty envelope. 

When compared with previous observations, the radio flux appears to vary.
In particular, the decrement of the 6~cm flux (from 1991 to 2002), does not agree with an expansion of the main
shell but appear to be related to the characteristics of the stellar object, whose fast evolution is difficult to be
explained in the framework of the recent
models of post-AGB  evolution.
  Further multi-frequency  monitoring and mapping with the ATCA fully
equipped with millimetre receivers are necessary to confirm such variation with a longer time baseline.
Up to now, variation (decrement) of the optically thin radio emission have been observed only in other two objects:
the proto-PN CRL 618 and the PN NGC 7027. In both cases, the observed variations have been interpreted in terms of evolution of the central object. The paucity of objects where such variations have been followed makes further radio observations of SAO~244567 very important.
as typical  timescale of such variations can be used to test current evolutionary models. 
%\section*{Acknowledgments}
%\end{document}

\end{document}